\documentclass[twocolumn,twoside]{svmultivs_gm} 
\usepackage{graphicx}
\usepackage{hyperref}
\hypersetup{
    colorlinks=true,
    linkcolor=blue,
    filecolor=magenta,      
    urlcolor=cyan,
    pdftitle={Overleaf Example},
    pdfpagemode=FullScreen,
    }
\title*{Large satellite constellations and their potential impact on VGOS operations}
\titlerunning{Satellite Megaconstellations and VGOS}
\author{Federico Di Vruno~$^1$, Vincenza Tornatore~$^2$}
\authorrunning{Di Vruno and Tornatore} 
\authoremails{vincenza.tornatore@polimi.it, Federico.DiVruno@skao.int}
\institute{1. SKA Observatory, Jodrell Bank, Manchester, United Kingdom\\ 
	2. Politecnico di Milano, DICA, Milano, Italy}
\ContactAuthorName{Vincenza Tornatore}
\ContactAuthorTelephone{+39 02 23996502}
\ContactAuthorEmail{vincenza.tornatore@polimi.it}
\NumberofInstitutions{2}
\InstitutionPostAddress{1}{SKA Observatory, Jodrell Bank Observatory, Macclesfield, Cheshire, SK11 9DL}
\InstitutionCountry{1}{United Kingdom}
\InstitutionWebPage{1}{https://www.skatelescope.org/ska-organisation/skahq/}

\InstitutionPostAddress{2}{Politecnico di Milano,DICA 
	(Sezione Geodesia e Geomatica),
	Piazza Leonardo da Vinci 32,
	I-20133 Milano}
\InstitutionCountry{2}{Italy}
\InstitutionWebPage{2}{https://www.polimi.it/en/}
\begin{document}  
\maketitle       
\abstract{}
Large LEO satellite constellations (or so-called Mega-constellations) will significantly change the view of the sky in some radio frequency bands. For VGOS telescopes it is important to understand the potential impact these constellations will have in their operations, what is the risk of its receivers going into non-linear behaviour and how much additional power would a telescope receive if observing in the same frequencies where satellites are transmitting. This work describes three of these new constellations (as they would look fully deployed) and summarizes the results of a particular study considering two VGOS telescopes (Onsala and Wettzell).

%
%
%
\section{Introduction}
The industrialization of spacecraft construction, and the lowering in costs of space launches has paved the way 
for big plans in Low Earth Orbit (LEO). Large satellite constellations like Starlink phase 1(with 4400 satellites) and OneWeb phase 1 (with 648 satellites) 
are already in the deployment phase, others like Project Kuiper (from Amazon) or Guowang (from China) are in their development phase and others with even larger numbers 
are being filed into the International Telecommunication Union (ITU) system (see Table \ref{NumConstell}). With altitudes between 500 km and 1200 km, these new constellations will surround the planet almost homogeneously. 
From a radio telescope point of view, the situation in the sky will change considerably. This change is already evident in the number of active satellites in LEO, from about 2000 in 2018, to more 
than 5000 in 2022, and the trend suggests it may reach hundred of thousands in this decade \cite{Lawrence}. 

Until now, most of the satellites for internet communication 
were located in the geostationary belt (at approximately 35780 km altitude), appearing fixed in the sky for a terrestrial observer \cite{Petrachenko_RFI}. The new LEO satellites will orbit the Earth with a period 
of about 90 minutes and will be seen as hundreds to thousands of bright and fast-moving radio sources in the sky with downlinks in frequency bands from 10.7 GHz up to 76 GHz (see Section \ref{Frequencies}).

Contrary to the situation with terrestrial radio frequency interference (RFI), it is not possible to build radio telescopes far away from satellite transmissions [1], 
the challenge is further increased due to the opposite pointing direction of radio telescopes and user downlink antenna beams. 

The typical power flux density ($PFD$) of satellite constellations is in the order of $-146 dBW/m^2$ (\cite{Starlink_ph1}, \cite{OneWeb_ph1}) in $4kHz$ or an equivalent to $62*10^6 Jy$, 
i.e. more than 7 orders of magnitude brighter than a typical VGOS source \cite{Petrachenko_WG3}. These strong signals will require a radio astronomy receiver to have a 
large dynamic range to accommodate the RFI and still be able to detect faint cosmic sources in other frequency channels within the receiver band. 
This is normally possible for modern radio astronomy receivers, but it can be different in some particular situations such as total power bolometric receivers or receivers with a low effective number of bits (ENB) \cite{Cooper_bits}.

\section{Large LEO Constellations}

Radio astronomy has been dealing with satellite transmissions since the very first satellites were launched back in the 1960s. Implementing different strategies such as using analog receivers with large dynamic ranges, smart scheduling, and RFI flagging among others, radio telescopes have been more or less able to mitigate (or avoid) the effect of these strong radio transmissions towards Earth \cite{RFI_Baan}. In conjunction with these strategies, spectrum management has also played a key role in dealing with the effects of satellites, several radio astronomy groups have worked at national, regional and international level for the protection of the radio astronomy service (RAS) frequency bands allocated by the International Telecommunication Union (ITU). Some with successful results, like the GLONASS example, and sometimes with battles that still ongoing 20 years after satellite deployment like in the IRIDIUM case \cite{Cohen}.
\begin{table}[htb!]
\caption{Some of the large LEO constellations in deployment and planned.}
\begin{tabular}{|l|c|c|} \hline
Constellation & Number of Satellites &  Altitude [km] \\
\hline
Starlink Phase 1 & 4,400 & 550  \\
OneWeb Phase 1 &   648 &  1200 \\
Amazon Phase 1 & 3,200 &  $\sim$ 600    \\
Guowang (GW) & 13,000 & 590 to 1145  \\
Starlink VLEO & 7600 & 340  \\
Telesat & 298 & 1000  \\
Starlink Phase 2 & 30,000 & 328 to 614 \\
OneWeb Phase 2 & 6,372 & 1200   \\
Cinnamon--937 & 327,320 & 550 to 643  \\
\hline
\end{tabular}
\label{NumConstell}
\end{table}
The exponential growth in the number of active satellites in Low Earth Orbit \cite{Lawrence} could result in  more than 2000 satellites above the local horizon at any moment in time. Radio telescopes are sensitive to any transmitter in line of sight through its main beam or antenna sidelobes.

\subsection{Walker-Delta constellations}
\begin{figure}[htb!]   
\includegraphics[width=.5\textwidth]{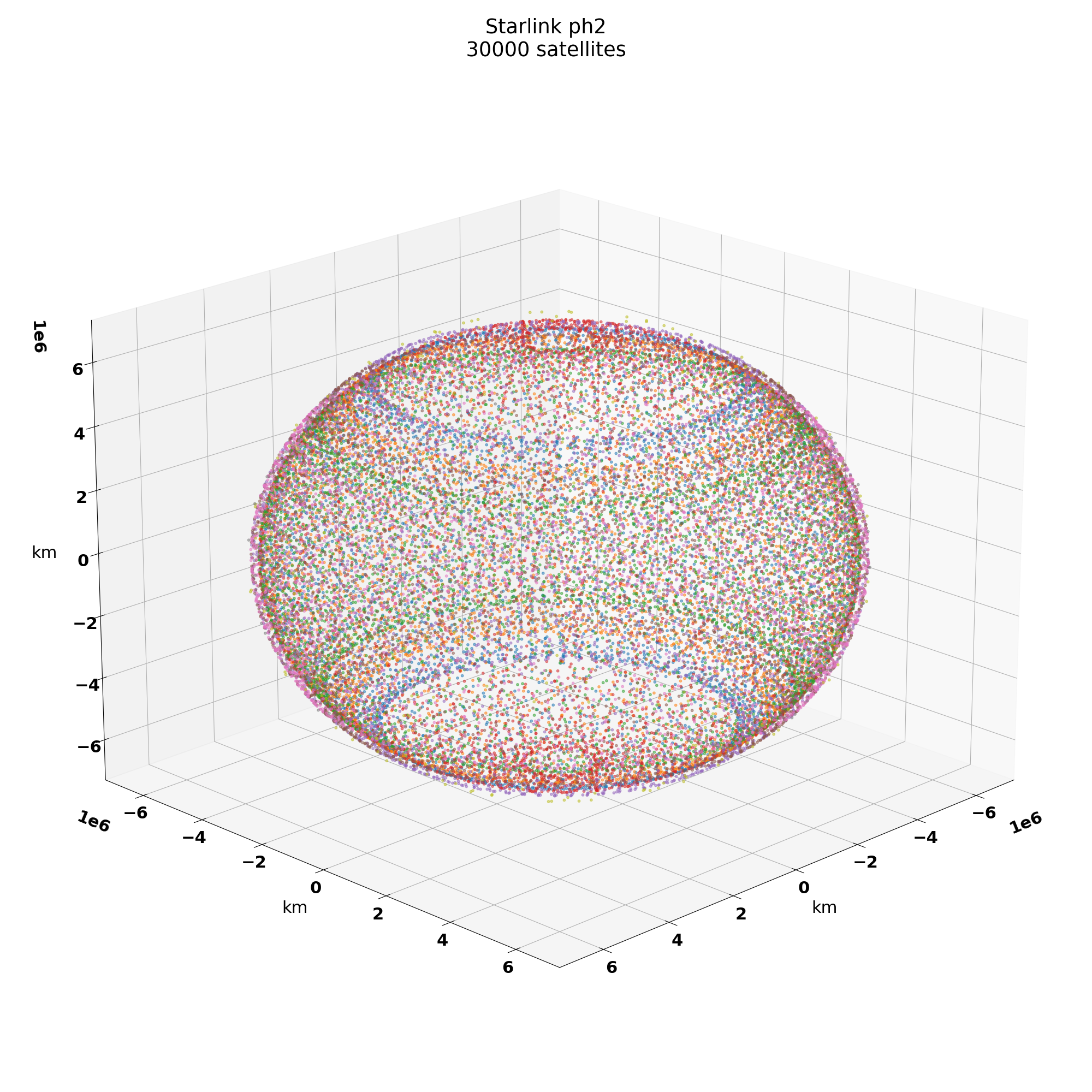}
\caption{View of Starlink Phase 2 constellation with 30000 satellites, different colors are used for each of the shells of the constellation Frequency 
bands used by some of the Satellite Constellations}
\label{StarlinkPh2}             
\end{figure}

All these new constellations follow a "Walker Delta" type of distribution, composed of orbital \textit{shells} at a certain altitude, each shell contains several \textit{orbital planes}, with a certain inclination with respect to the Equator and distributed homogeneously in the 360 degrees of 
right ascension. Each one of the constellation's planes contains $N$ satellites, a representation of Starlink Phase 2 can be found in Figure \ref{StarlinkPh2}.

A shell of a Walker-Delta constellation \cite{Walker} is described by $i = t/p/f $ where $i$ is the inclination, $t$ is the total number of satellites, $p$ is the number of equally spaced planes, and $f$ is the relative spacing between satellites in adjacent planes. This description makes it very simple to simulate any of these constellations with the purpose of studying its geometric distribution 
in LEO and also its effect on radio telescopes. It is also possible to use existing Two-Line Elements (TLEs) to obtain the approximate position of existing 
satellites in space, which can be useful to compare observations to simulation.

\begin{figure*}[htb!]
\includegraphics[width=5.0cm]{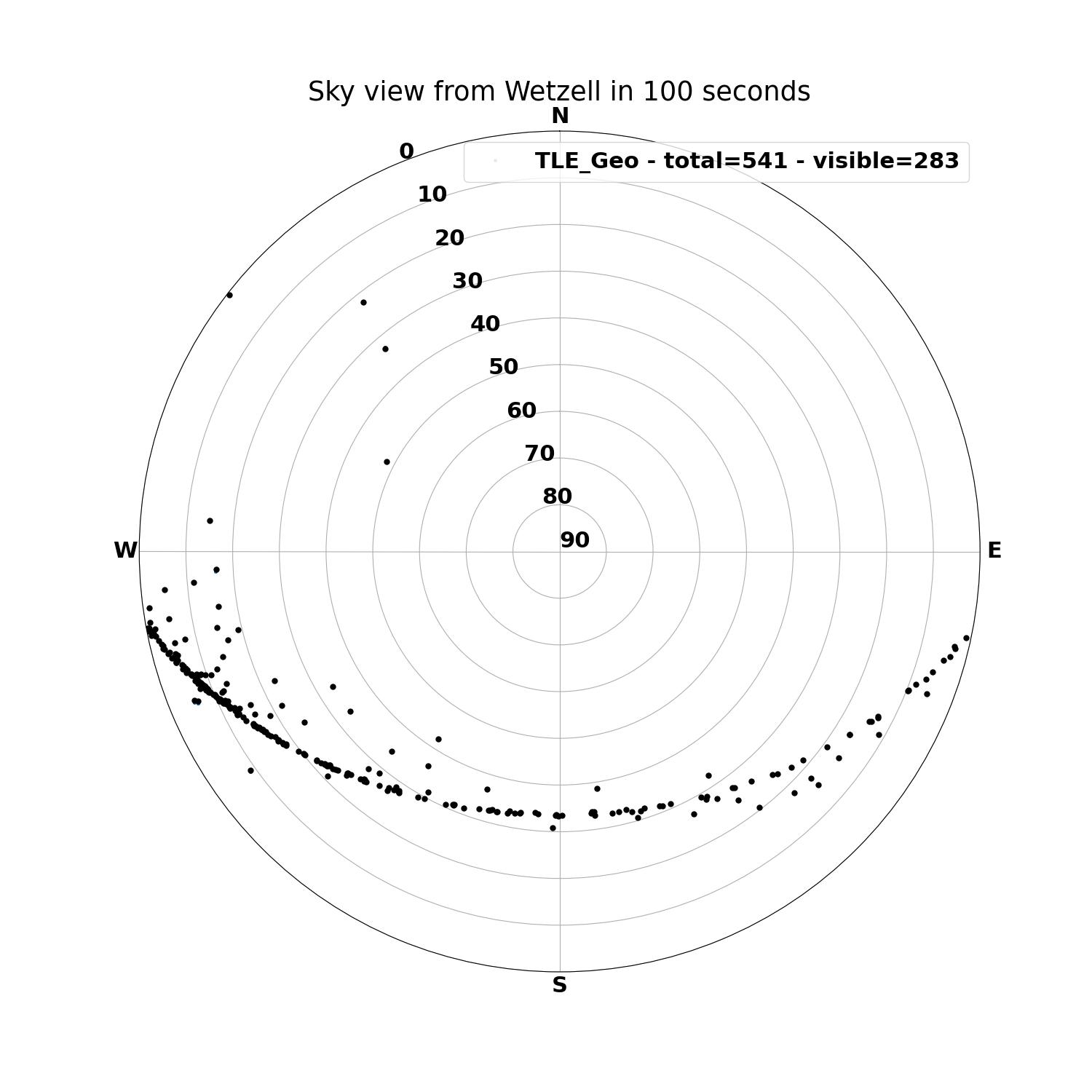}
\includegraphics[width=5.0cm]{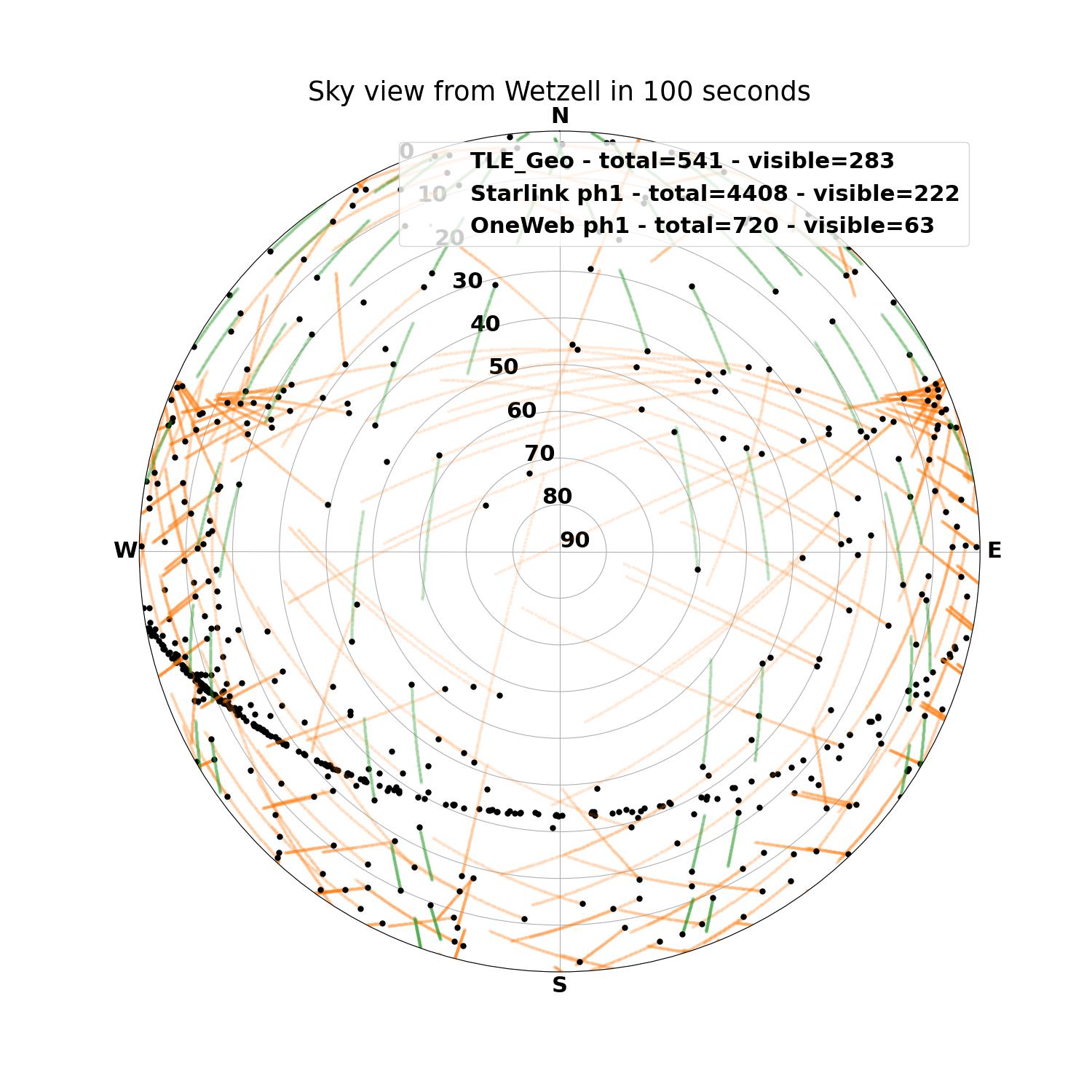}
\includegraphics[width=5.0cm]{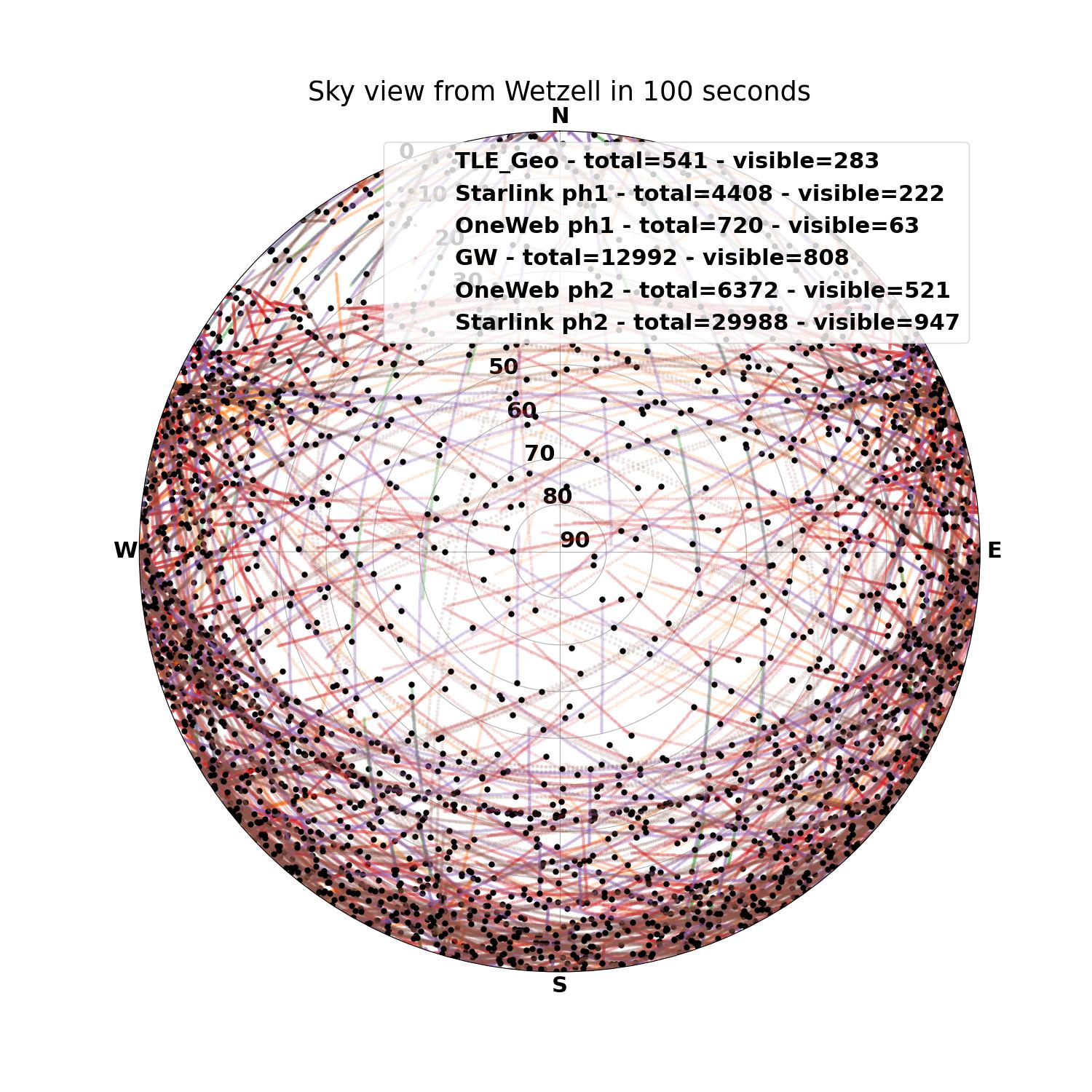}
\caption{Sky view from the Wettzell VGOS station with only Geostationary satellites (left), simulation of SL1 and OW1 constellations fully deployed (middle) and simulation of 6 large LEO constellations fully deployed (right). The term "visible" is used for satellites above the horizon, as radio telescopes can detect satellites in any direction in the sky.}
\label{WzSat}
\end{figure*}

Figure \ref{WzSat} shows a qualitative view of the sky from the Wettzell VGOS station (lat 49 degrees), with the position of different satellite constellations simulated for 100 seconds. It is simple to see how the density of satellites in the sky will drastically change in the near future if all constellations planned are deployed.

\subsection{Radio frequencies} \label{Frequencies}
Satellite constellations transmit their downlink signals in frequencies allocated to the Fixed Satellite Service (FSS). Table \ref{FreqBand} contains some of the currently in-use and planned FSS bands and it is important to note the proximity to some ITU protected RAS bands immediately adjacent  
or in very close proximity. 
\begin{table}[htb!]
\caption{Frequency bands used by some of the Satellite Constellations.}
\begin{tabular}{|l|c|c|} \hline
Frequency & Band name &  Protected RAS bands \\
          &      &   (primary)  \\
\hline 
10.7 - 12.75 GHz & Ku &  10.6 - 10.7 GHz  \\
19.7 - 20.2 GHz &   Ka &   22.21 - 22.5 GHz \\
37.5 - 42.5 GHz & V &   42.5 - 43.5 GHz   \\
71.0 - 76.0 GHz & E &  76 - 77.5 GHz  \\
\hline
\end{tabular}
\label{FreqBand}
\end{table}
The close vicinity of the satellite's downlinks to radio astronomy bands is a matter of concern for radio astronomers and spectrum managers. As an example, the protection of the 10.6-10.7 GHz Radio Astronomy Service (RAS) band, which includes a \textit{passive band} in 10.68-10.7 GHz protected by the footnote RR No. 5.340 in the ITU-R Radio Regulations (RR), was studied for the Starlink Ph1 and OneWeb ph1 constellations in \cite{ECC271}, with the conclusion that both systems should not use the first 250 MHz channel to protect the RAS band. These signals can not only impact sensitive observations in the RAS protected bands, but can also affect wideband receivers which include the frequency range of user downlinks. Such wideband receivers (from 1 to 14 GHz in the case of VGOS) are necessary to conduct cutting edge science or Geodesy \cite{Petrachenko_WG3}. 

This paper focuses on the downlink frequency range 10.7 to 12.75 GHz where both OneWeb and Starlink have divided the band in 8 channels of 250 MHz each. The study can be replicated for higher frequency bands with the appropriate modification of satellite and telescope characteristics.

\section{Potential impact on VGOS}
%
By using large reflector antennas pointed towards the sky and wideband receivers covering the frequency range 1 to 14 GHz \cite{Petrachenko_WG3}, 
VGOS telescopes can be impacted by downlinks of the large satellite constellations in different ways. In fact the VGOS bandwidth is wide while the protected Radio astronomy band is very narrow in and Starlink and OneWeb frequencies use a considerable portion of spectrum. The severity of this 
impact depends on the interaction between the radio telescope beam and the satellite downlink beams. One of the most important aspects is how much 
a correlated baseline can be affected, as the primary product of a VGOS observation. Nevertheless, the multi-dimensionality of this problem requires an analysis of the 
complete signal reception mechanisms and how each part of the signal chain may be impacted.

In a typical VGOS schedule, targets are observed with durations in the order of seconds to tens of seconds, the position of the target in the local sky and the density of satellites deployed will define how much interference will be seen by the telescope. The instantaneous received power from all  satellites above the horizon
may saturate the analog signal chain (low noise amplifiers, mixers, etc), causing non-linearities that would render the complete receiver band unusable, even if the 
digitizer band is tuned to a completely different frequency than the satellite downlinks channels. If the RFI power is not as strong and the analog signal chain remains linear, then there
can be two possible scenarios: 
\begin{itemize}
\item First scenario: when the observed band is outside of the satellite downlink frequency range, in which case out of band emissions from the satellites could be 
a problem depending on their level. This work is not focusing on this, but \cite{ECC271} has studied that case.

\item Second scenario: if the observing band falls within one satellite downlink band (250 MHz channels) or vice versa, strong RFI will be received by the VGOS antenna. This RFI can potentially be mitigated by correlation as long as the number of bits in the digitizer are enough to correctly digitize the signal. Since a VGOS digitizer has only two bits, the total integrated 
RFI needs to be lower (practically at least 10 dB lower or $1/10$) than the integrated noise power of the receiver \cite{Cooper_bits}.
\end{itemize}

Non-linearities and lack of headroom for RFI are transient phenomena and can be considered in terms of a data-loss associated with the moments where one satellite is going through the main beam of the radio telescope. The issue of out of band emission is related to long integrations 
and needs a comparison between the level of integrated RFI vs the integrated level of the astronomical source under observation. The following section describes a simulation method and presents a particular case for the Starlink phase 1, OneWeb phase 1 and Starlink phase 2 constellations to estimate data loss due to strong received power and the total aggregated RFI, the effects of the correlation is not included in this work as is currently under study by the authors.

\section{Simulation methodology}
The simulation is based on the Equivalent Power Flux Density ($epfd$) concept (see \cite{S.1586}), where the satellite constellation is propagated for a defined time duration, obtaining the coordinates and attitude of every satellite for each time step. Then, the telescope antenna is pointed towards a defined {\itshape sky-cell} in azimuth and elevation and for each of the simulated time steps, the received power from all satellites above the horizon is calculated with the formula:
\begin{equation}
\label{eq:SatPow}
P_{rx_{(t,p)}}=\sum_{i=0}^{N_{sat}}(PFD_{sat_{(i,t)}} * A_{eff_{RAS_{(i,t,p)}}})
\end{equation}
where:\\
$t$ = time step \\
$p$ = pointing direction \\
$i$ = satellite index \\
$PFD_{sat}$ = Satellite power flux density in $W/m^2$ towards the telescope location\\
$A_{eff_{RAS}}$ = Effective area of the telescope antenna in $m^2$ towards the satellite position
\begin{figure}[htb!]
\includegraphics[width=7.5cm]{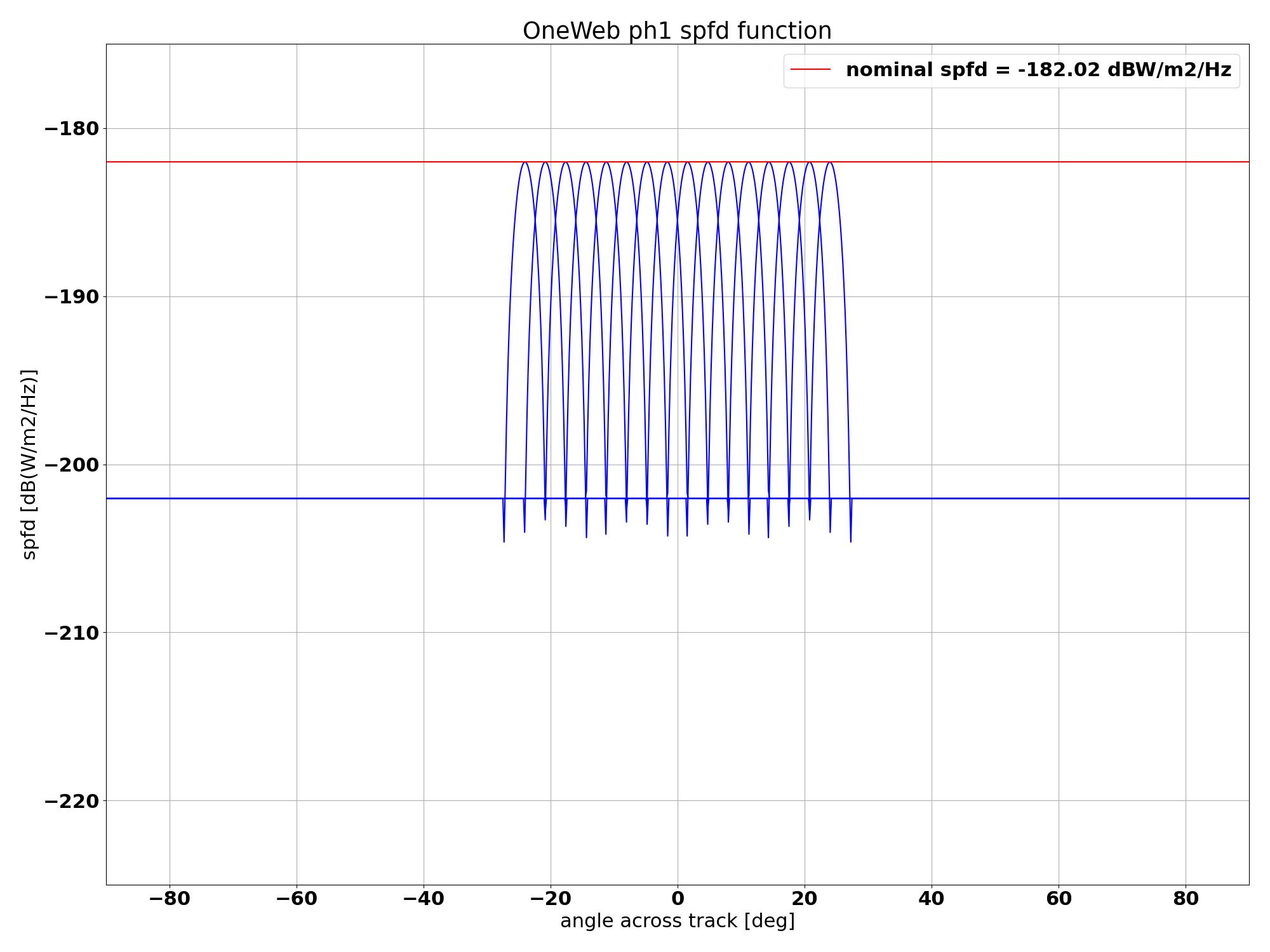}
\includegraphics[width=7.5cm]{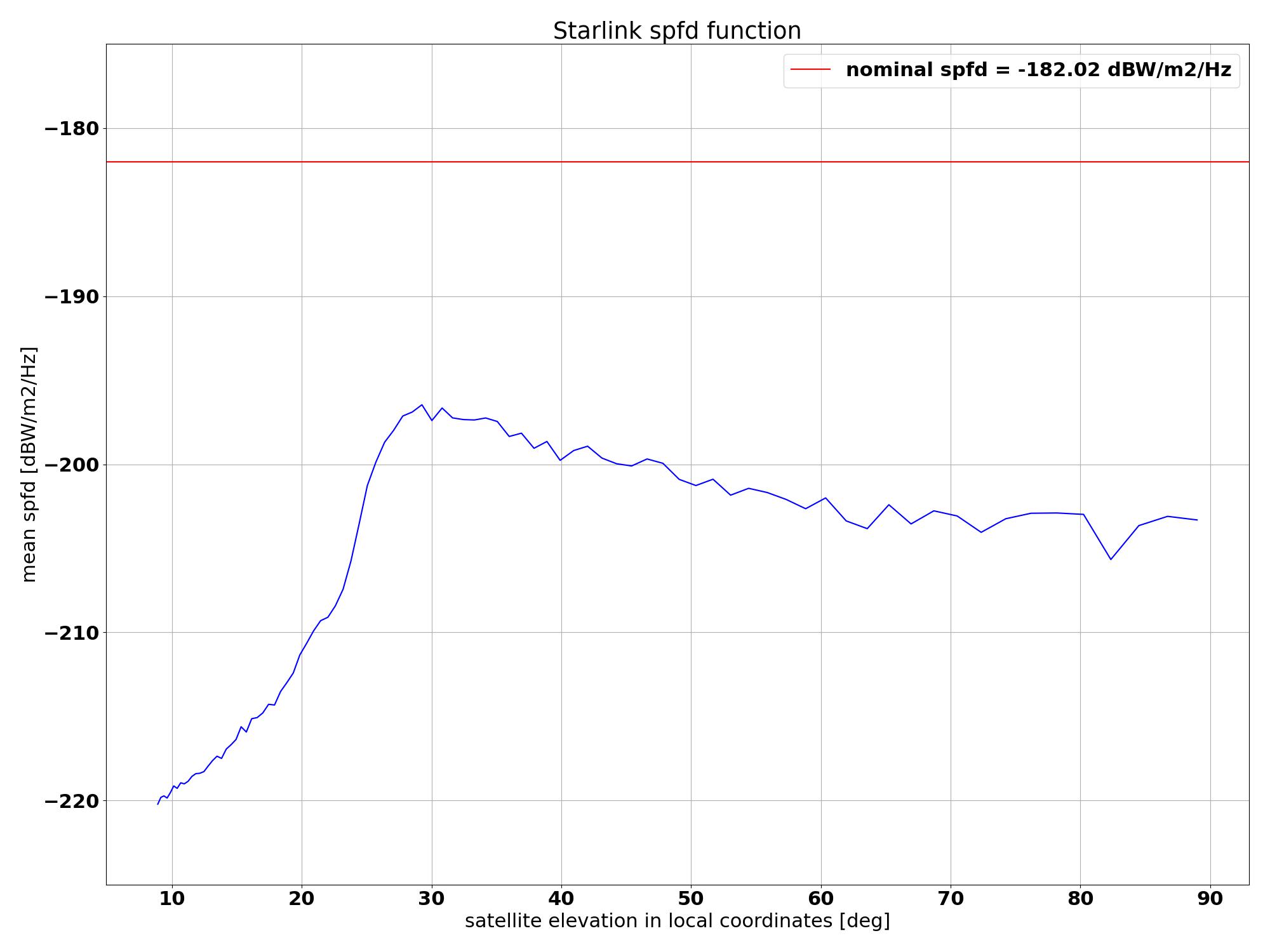}
\caption{OneWeb $SPFD$ model(top), Starlink $SPFD$ model(bottom), the red line marks the maximum $SPFD$ level of $-182 dB/m^2/Hz$.}
\label{beams}
\end{figure}

This calculation is iterated for a number of {\itshape trials} (typically hundreds to thousands), where each try has a random start time of the constellation and therefore contributes to a statistically representative result.
In situations where multiple frequencies are 
calculated, like for example the case of OneWeb with its 16 fixed-beams antenna (see Figure \ref{beams}), the number of channels is added to the result. 
Therefore the final calculation results in a data cube with four dimensions, namely number of iterations, 
number of pointing directions, number of time steps, and number of channels:
$N_{iters}$, $N_{pointing}$, $N_{time}$ and $N_{channel}$.

Although the original $epfd$ calculation as defined by the ITU uses telescope pointings in local coordinates (Alt,Az), this work considers pointings in celestial coordinates (Ra,Dec) as this allows to understand how celestial positions in different declinations can be impacted by satellite constellations transmissions.


\subsection{Satellite position propagation}
Using the Python package Cysgp4 \cite{Cysgp4} and the Astropy Coordinates package \cite{Astropy}, the position of the satellites in horizontal coordinates (Alt,Az) and Sky coordinates (Ra,Dec) are calculated for each timestep and each iteration (see Figure \ref{horizontal_view}).

\subsection{Satellite power flux density $(PFD)$}
The $PFD$ from each satellite in a constellation is modelled based on publicly available information (ITU documents and FCC filings). To calculate the power flux density towards the telescope site, the coordinates of the telescope in the satellite reference 
frame are also calculated using the Python package cysgp4 \cite{Cysgp4}. 

OneWeb satellites are modelled based on the information available in the ECC report 271 \cite{ECC271}, 
with 8 channels in the $10.7-12.75 ~ GHz$. A fixed beam antenna pattern, like the OneWeb system, makes it simpler to 
calculate the received power in a deterministic way.

The $PFD$ from Starlink satellites is more complex to model since they 
have an antenna array that can produce, and electronically steer, several beams in one or multiple frequency channels. The mean $PFD$ from a Starlink satellite is modelled as a function of the elevation of the satellite, obtained from a 
Monte Carlo simulation in where the steering angle, the number of beams and the position of satellite and observer was varied a large number of times. Starlink satellites are modeled as one 
frequency channel at a time.

\subsection{Radio Telescope antenna}
 
The radio telescope antenna is modelled based on \cite{RA.1631}. While this model is not a real measurement of the antenna pattern of a radio telescope, it is based on real measurements and is considered 
as a worst case for compatibility studies. To obtain the gain towards 
the satellite, the angle between the pointing direction and the position of the satellite is calculated.

The Effective Area of the antenna is calculated with the following equation: \begin{equation}
\label{eq:A_Eff}
A_{eff} = G_{RAS}*(\lambda^2/(4*\pi)) 
\end{equation}

\begin{figure}[h!]
\includegraphics[width=.5\textwidth]{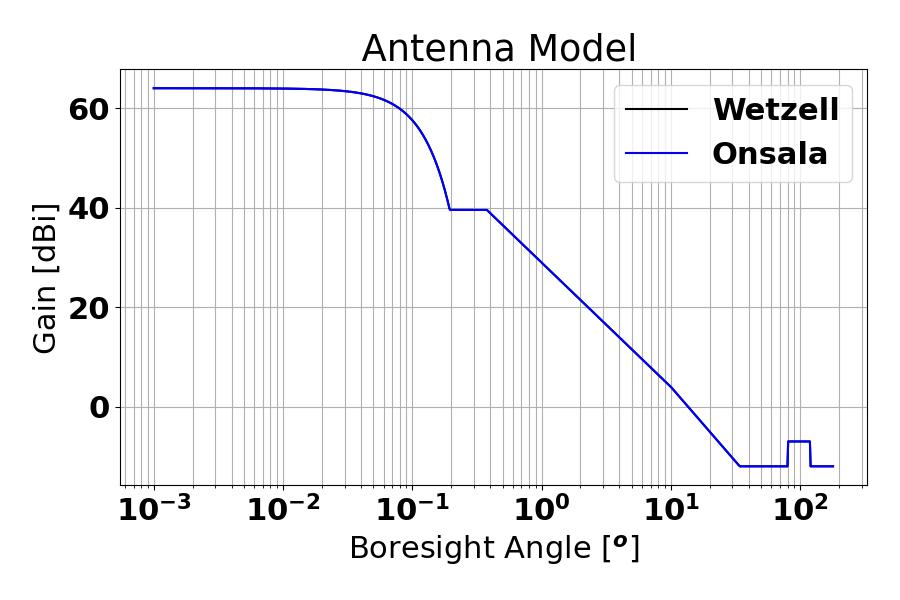}
\caption{Antenna pattern as defined in ITU-R RA.1631.}
\label{RAS_antenna}
\end{figure}

\subsection{Correlation}

Interferometry can greatly mitigate the effects of RFI, especially when the baselines are large like in the case of VLBI \cite{Thompson_RFI}. Although Thompson and others have studied the effect that long baselines have over single RFI transmitters (and stationary), the situation is not the same when potentially hundreds of transmitters using the same frequency and bandwidth are received simultaneously as can happen now.

For example in \cite{Petrachenko_RFI}, Petrachenko identifies the $10.7-12.75~GHz$ range as a usable frequency range as only Geostationary satellites were using that frequency at that time.
Now the received RFI signal at one antenna will be the sum of the signals from all satellites above the horizon (of course with different levels of attenuation). This analysis is deferred to a further update of this work.

\subsection{Saturation Limit threshold}

Digital processing operations in a radio telescope can be applied as long as the analog and digital signal chains behave in a linear manner; 
strong enough signals will generate non-linearities corrupting the complete receiver band for the duration of the interference. Defining the level where 
a receiver goes non-linear is not a simple task and will depend on each particular receiver. In the case of the VGOS receivers a conservative 
value for total power of $-50~dBm$ is considered to keep the analog signal chain within the linear regime. 

If the received power is below this linearity threshold, the analog signal can then be correctly digitized with a bandwidth of $1~ GHz$. Two scenarios can be identified: 
\begin{enumerate}
\item Digitizing a frequency range outside of the $10.7-12.75~GHz$, which should not have any complications since the signal chain behaves in a linear way and therefore this case will not be further studied;
\item Digitizing in a frequency range within the $10.7-12.75~GHz$. In this case is interesting to understand when the RFI produces a significant amount of power compared to the RMS noise of the receiver.
\end{enumerate}
Given the distinct characteristic of VGOS systems using a 2 bit correlator, it is reasonable to consider that there is not much headroom in the digital 
signal chain to accommodate for RFI, this work considers that any signal above or equal to the receiver's noise power will result in a data loss. This defines the second threshold as a spectral power flux density equal to the RMS noise of a $20~K$ receiver system ($-215~dBW/Hz$).

These two thresholds are used in the simulation; a first set of flags is produced when the total integrated power (considering the 8 channels of $250~MHz$ 
for each constellation) is higher than $-50~dBm$ (representing a total data loss) and the second one representing a data loss in the case of observing in the same frequency range as the satellite transmissions.

After these two flagging stages, low level RFI will still be present, it is of interest to understand how this will affect the correlation of the baseline. This will be further study in a future update to this work and compared to the thresholds defined in RA.769 \cite{RA.769}.

\subsection{Metrics} \label{Metrics}
Based on the threshold limits defined in the previous section, the following metrics are used:
\begin{enumerate}
\item Full Band Data Loss (\textbf{FBDL}): percentage of time that the complete band is lost due to very strong RFI, where the total received power is $>-50~dBm$;
\item Digitizer Data Loss (\textbf{DDL}): Percentage of the total observation time (single run multiplied by the number of iterations) that the instantaneous power 
spectral density is above 10\% of the integrated noise power of the receiver. This can be calculated as a function of the declination of the source;
\item Average Equivalent Spectral Power Flux Density (\textbf{aESPFD}): average value of the equivalent Spectral Power Flux Density during the observation 
time in each antenna. The eSPFD is calculated as the received spectral power flux density $[W/m^2/Hz]$ divided by the maximum effective antenna area, and it is useful to compare to the SPFD (in units of $Jy$) of a celestial source 
in the main beam of the antenna;
\end{enumerate}

\section{Case study simulation}
A specific study case was selected to understand the impact from several satellite constellations on two telescopes normally involved in VGOS observations, it is the intent to further expand this work into how correlation over the long baseline mitigates the RFI. 
The VGOS stations in Sweden (Onsala Observatory) and Germany (Wettzell Observatory) were selected as the test stations, using the parameters in Table \ref{VGOS_STAparam}, and Starlink phase 1, OneWeb phase 1 and Starlink phase 2 as constellations see Table \ref{ParSat}. The simulated observations were runned for 100 seconds in 1 second timesteps with 100 iterations.

\begin{table}[htb!]
\caption{VGOS stations parameters used for the simulation}
\begin{tabular}{|l|c|c|} \hline
Station & Wettzell &  Onsala \\
\hline
Location (lon,lat) (deg) & (12.88, 49.14)& (11.92, 57.39)  \\
Height (m)&   600 &  20 \\
Antenna Diameter (m)& 13 &  13 \\
Antenna Efficiency (\%)& 80 &  80 \\
Receiver bandwidth (MHz) & 1000 & 1000 \\
System Temperature (K) & 20 & 20  \\
ITU-R RA.769 & -240 & - 240  \\
threshold ($dBW/m^2/Hz$) &  &   \\
\hline
\end{tabular}
\label{VGOS_STAparam}
\end{table}

\begin{table}[htb!]
\caption{Parameters of satellite constellations used for the study \cite{Starlink_ph1}\cite{Starlink_ph2}\cite{OneWeb_ph1}.}
\begin{tabular}{|l|c|c|c|c|} \hline
Constellation & Altitude &  Inclination &  Number  & Satellites \\
              &          &              &of planes & per plane \\
\hline
Starlink ph 1 & 550 & 53 & 72 & 22 \\
\cline{2-5}
&  540 & 53.2 & 72 & 22 \\
\cline{2-5}
&  570 & 70 & 36 & 20 \\
\cline{2-5}
&  560 & 97.6 & 6 & 58 \\
\cline{2-5}
&  560 & 97.6 & 4 & 43 \\
\hline
Oneweb ph 1 & 1200 & 87.9 & 18 & 40 \\
\hline
Starlink ph 2 & 340 & 53 & 48 & 110 \\
\cline{2-5}
&  345 & 46 & 48 & 110 \\
\cline{2-5}
&  350 & 38 & 48 & 110 \\
\cline{2-5}
&  360 & 96.9 & 30 & 120 \\
\cline{2-5}
&  525 & 53 & 28 & 120 \\
\cline{2-5}
&  530 & 43 & 28 & 120 \\
\cline{2-5}
&  535 & 33 & 28 & 120 \\
\cline{2-5}
&  604 & 148 & 12 & 12 \\
\cline{2-5}
&  614 & 115.7 & 18 & 18 \\
\hline
\end{tabular}
\label{ParSat}
\end{table}

\begin{figure}[htb!]
\includegraphics[width=0.5\textwidth]{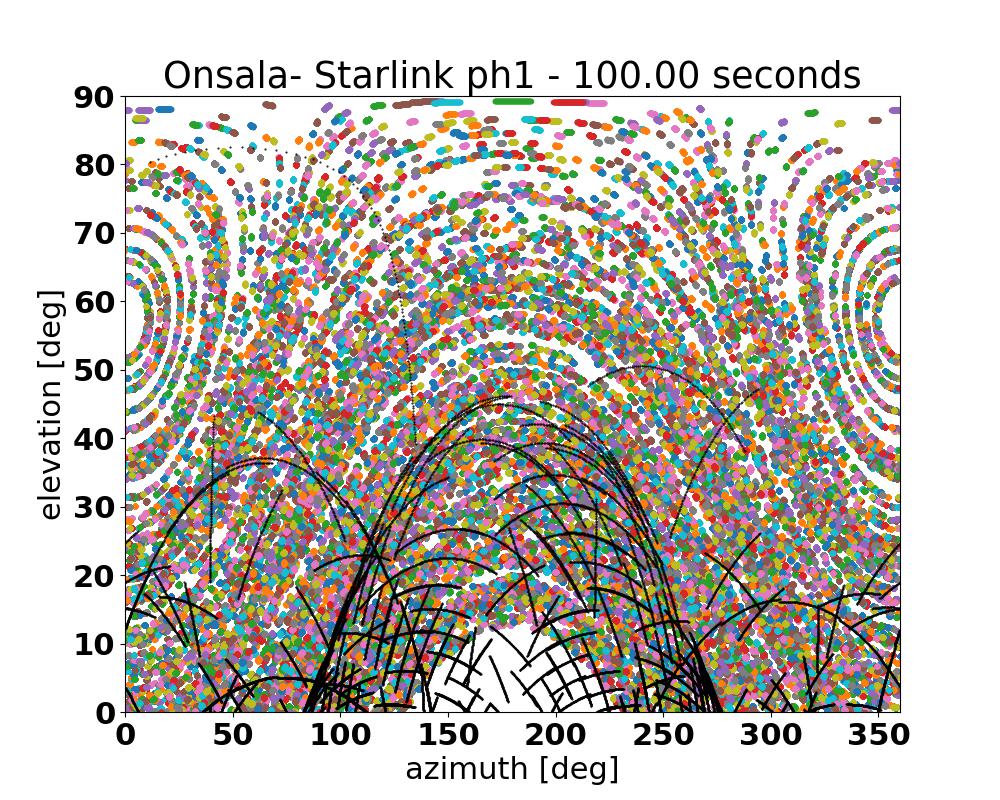}
\caption{Horizontal (Alt,Az) view of the pointing directions (in colours) and movement of the Starlink Phase 1 satellites (black) as seen from the Onsala Observatory for a time duration of 100 seconds.}
\label{horizontal_view}
\end{figure}

\begin{figure}[htb!]
\includegraphics[width=.5\textwidth]{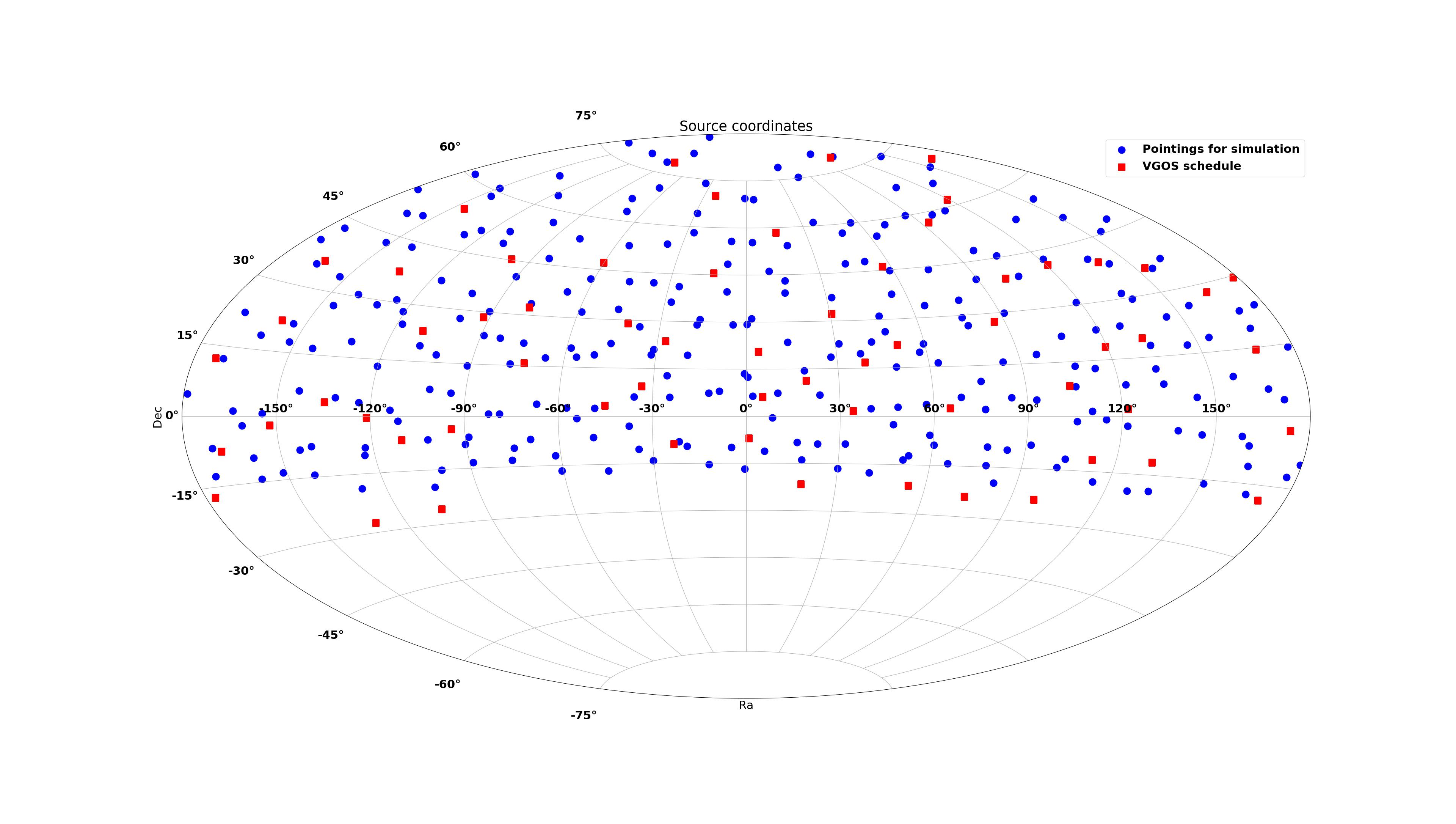}
\caption{VGOS schedule (2022Jan27) sources in red boxes, selected telescopes pointings for the simulation in blue circles.}
\label{Sources}
\end{figure}

Originally it was intended to use a real VGOS schedule, using real Ra, Dec of sources observed, but to get a more representative results of the impact as a 
function of source declination the number of sources was increased artificially to 277 in a random fashion, see Figure \ref{Sources} for a plot of the sources distribution. Figure \ref{horizontal_view} shows the view of the local sky in (Alt,Az) and how the celestial sources and the satellite constellation (in this case Starlink Phase1) move across the sky in that timeframe.

\section{Results} \label{Results}
The results for each one of the selected metrics is summarized here for each constellation simulated. 

\subsection{Full Band Data Loss (\textbf{FBDL})}
Notably, the analog saturation threshold was not reached due to the combination of maximum PFD from the satellites ($-98~dBW/m^2$ in $250 ~ MHz$) and 

\begin{figure*}[htb!]
\includegraphics[width=15cm]{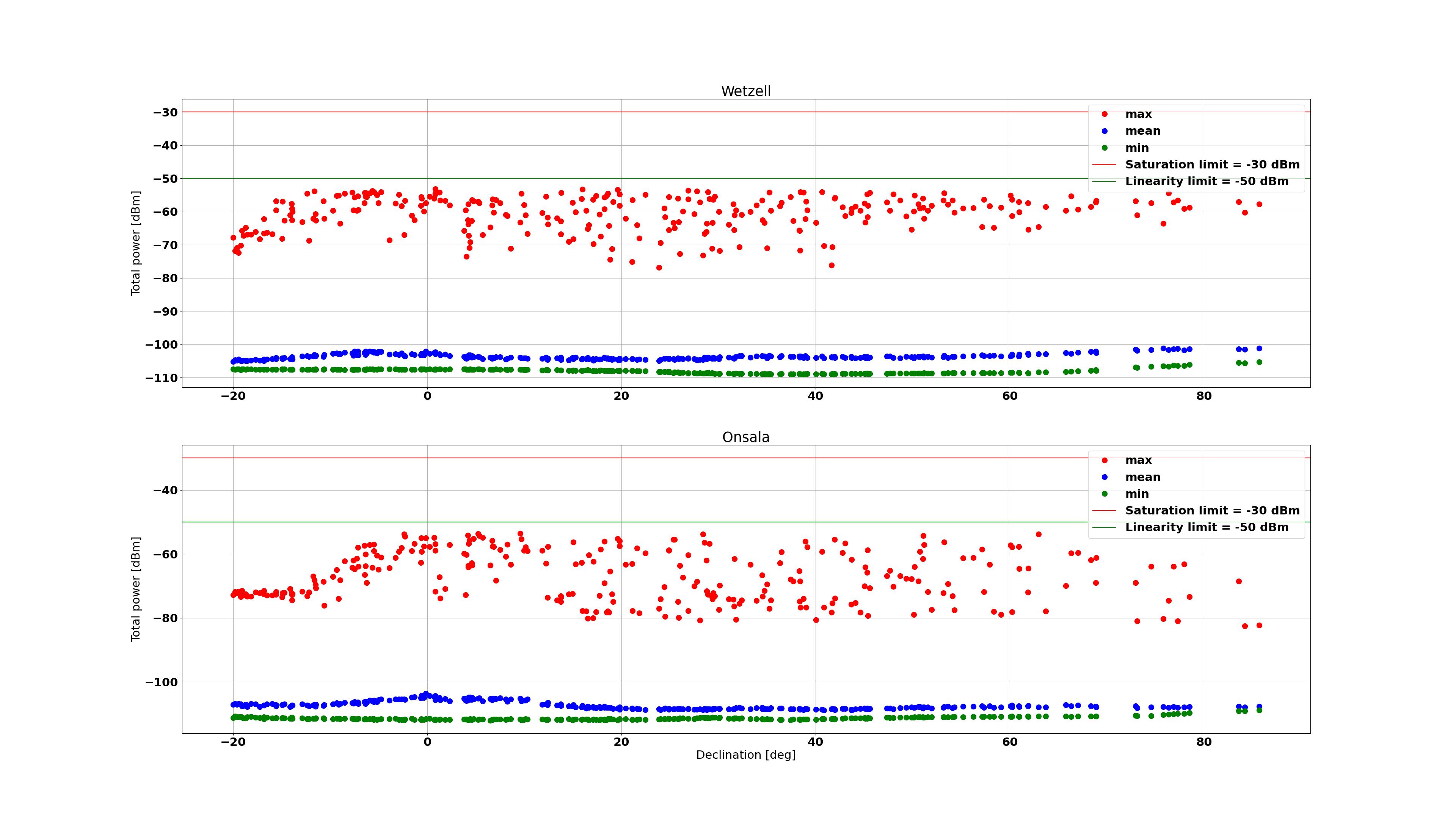}
\caption{Instantaneous power received by both VGOS antennas as a function of pointing declination with Starlink  phase 2 constellation. The Linearity threshold of $-50~dBm$ was not reached in any situation.}
\label{FBDL}
\end{figure*}

maximum effective area of the VGOS antennas ($106~m^2$ or $20.3~dBm^2$), as can be seen in Figure \ref{FBDL}. 
This shows that even with large constellations such as Starlink phase 2 the analog receivers would still behave in a linear fashion.

\subsection{Digital Data Loss (\textbf{DDL})} \label{Result_DDL}
\begin{figure}[htb!]
\includegraphics[width=7.80cm]{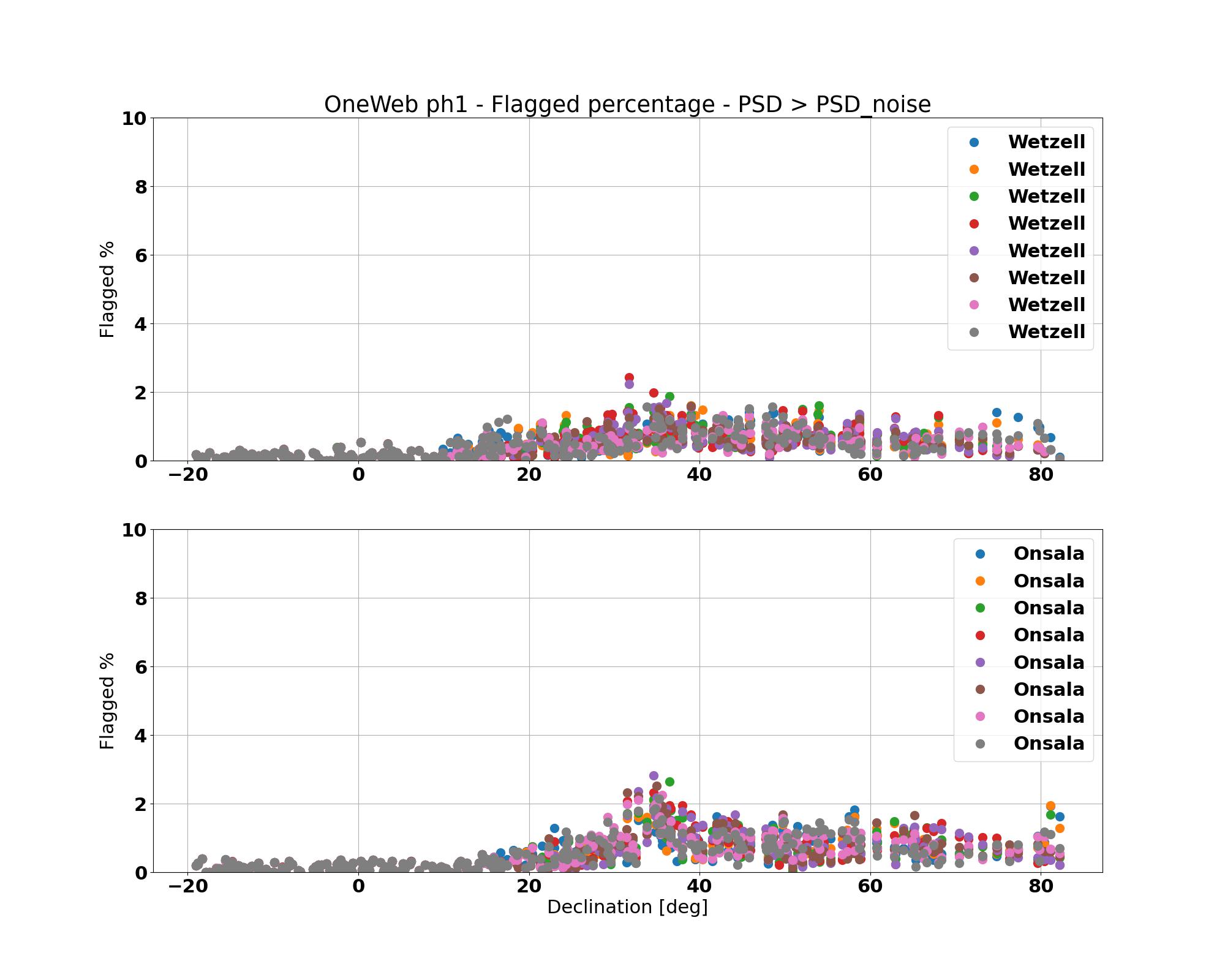}
\includegraphics[width=7.80cm]{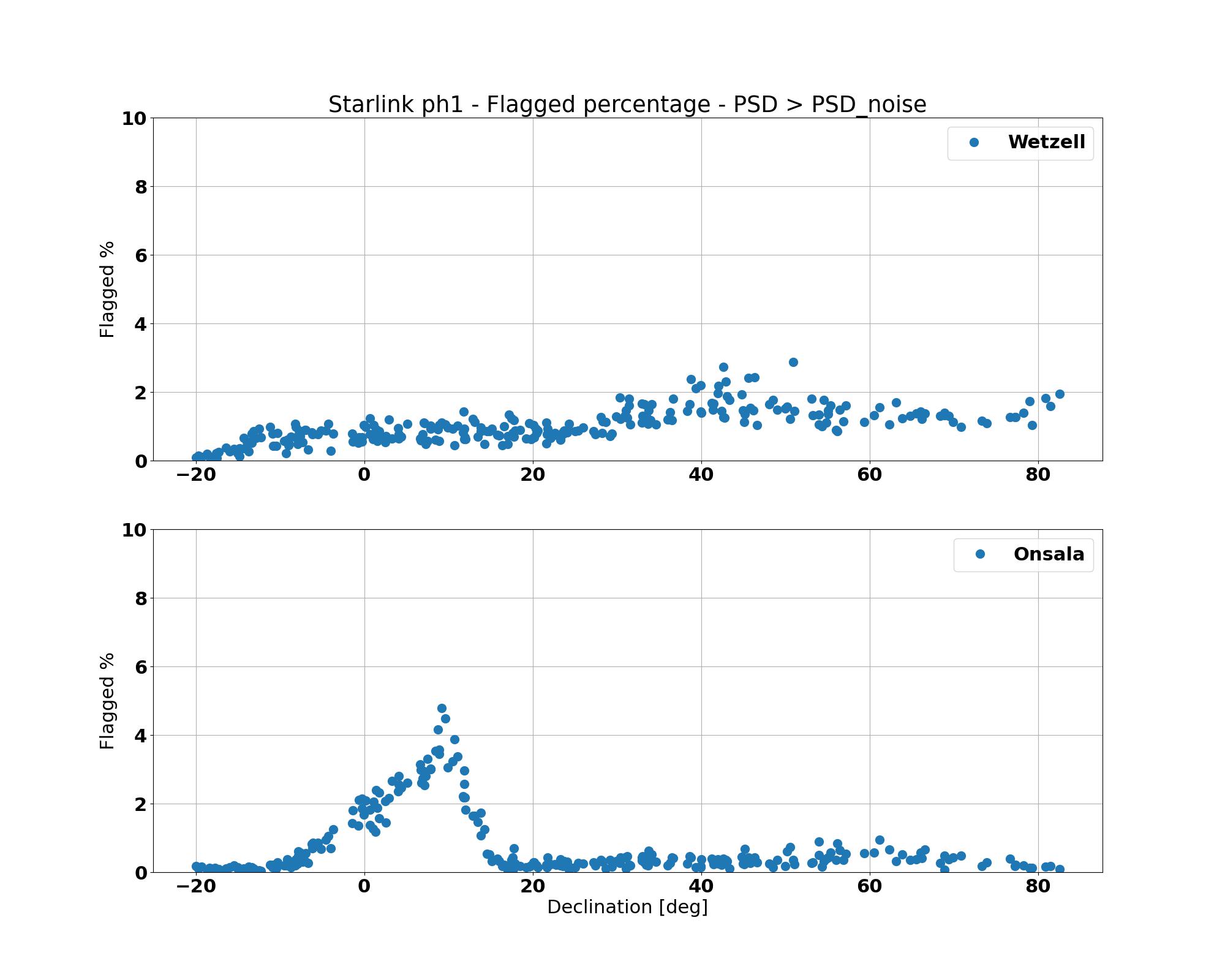}
\includegraphics[width=7.8cm]{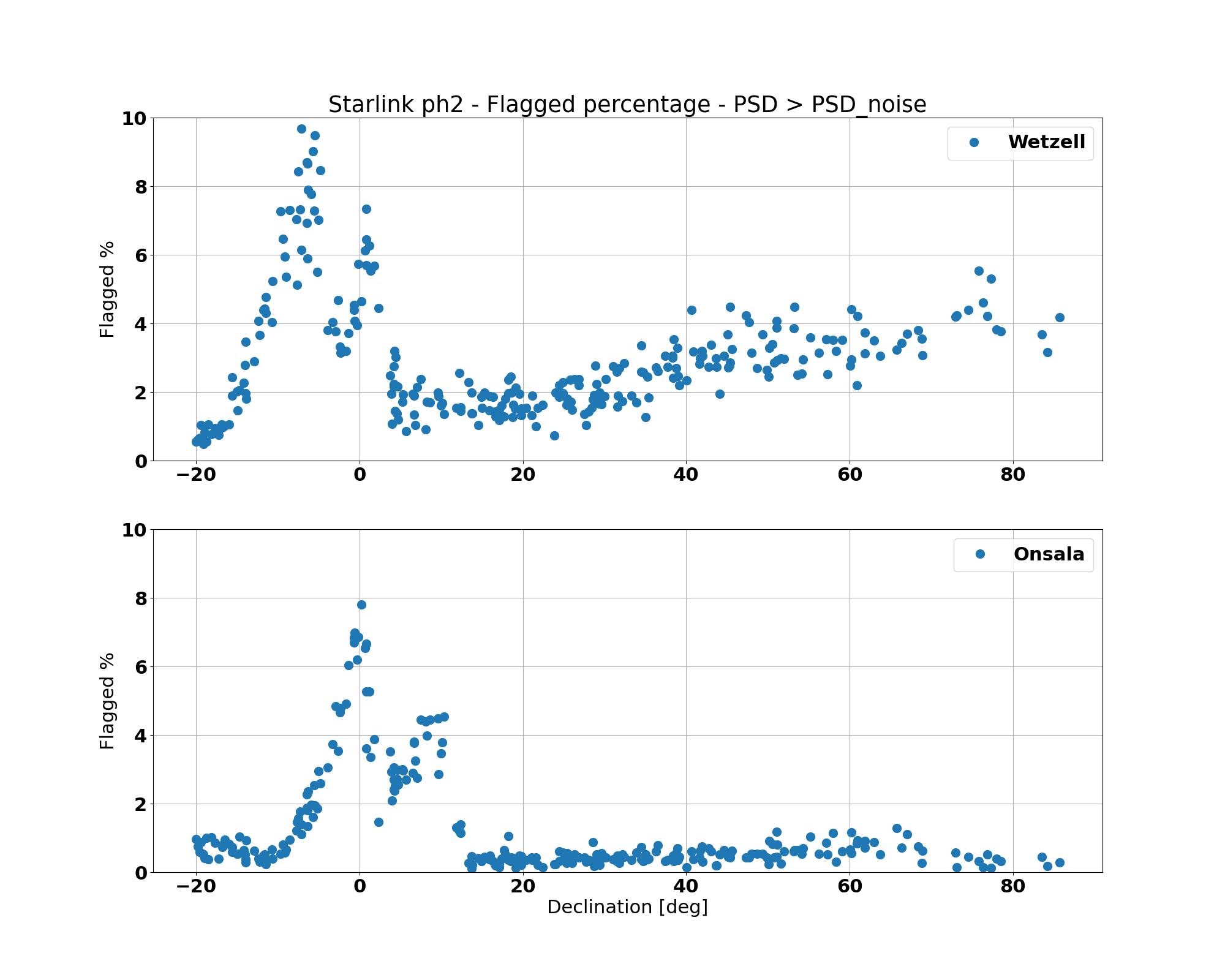}
\caption{Flagged percentage for each antenna and each constellation, a flag is raised when the power spectral density received is above the noise spectral density.}
\label{DDL1}
\end{figure}

When considering an observation coinciding in frequency with the downlinks of satellites (i.e. in within the $10.7-12.75~GHz$) the DDL varies as a function of declination of the observed source and observatory latitude. This effect is attributable to the different structures of each constellation's density of satellites around the Earth and the latitude of the observer. This shows that impact to VGOS stations (and radio telescopes in general) will strongly depend on the observatory latitude. See Figure \ref{DDL1}.

\subsection{Average Equivalent Spectral Power Flux Density (\textbf{aESPFD})}
After a certain percentage of the observed data was lost as DDL (see section \ref{Result_DDL}, the aESPFD is calculated for each constellation as a function of declination. In this case the flagged percentage is calculated as the product of the flags from the previous section for each antenna.

\begin{figure}[htb!]
\includegraphics[width=7.80cm]{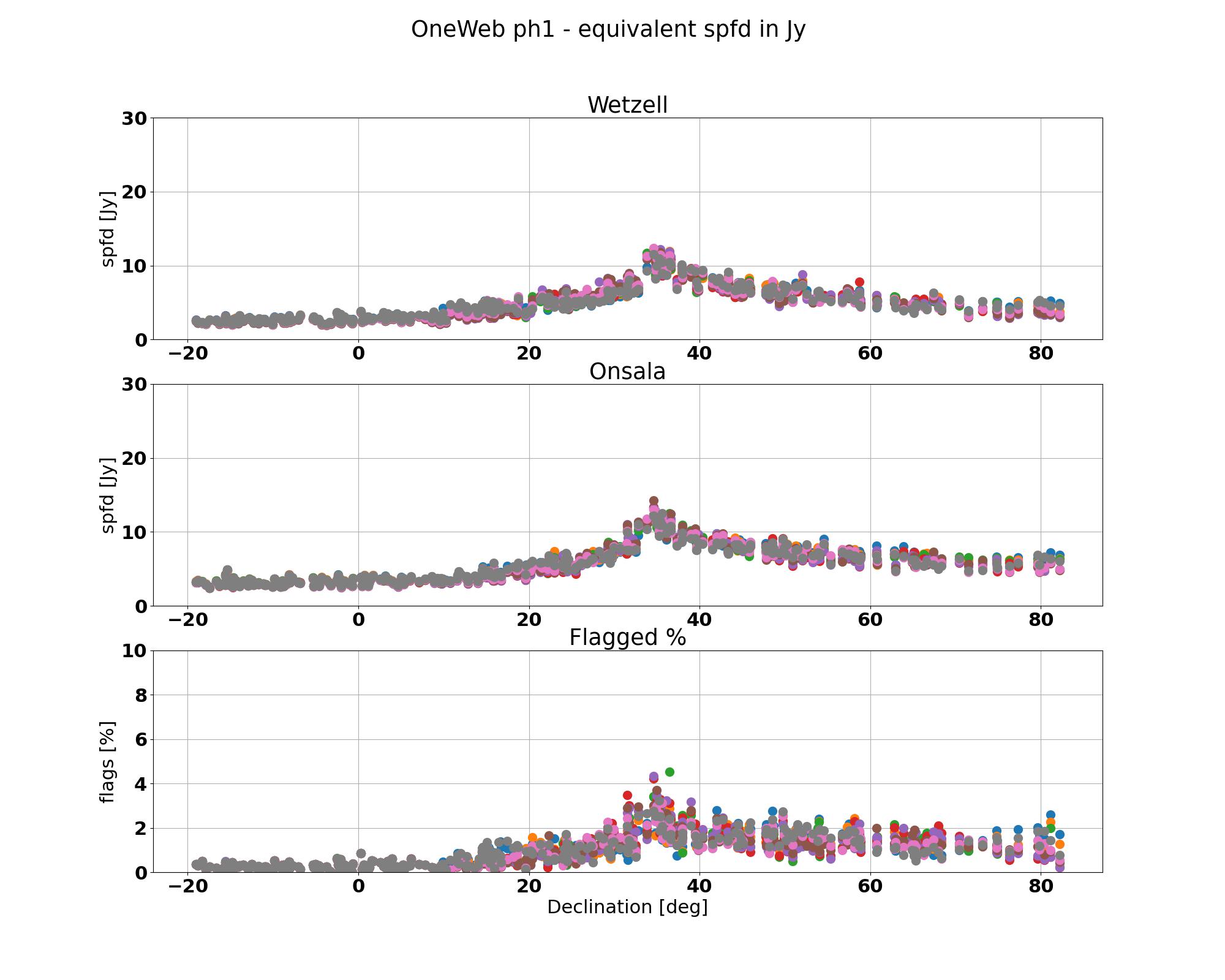}
\includegraphics[width=7.80cm]{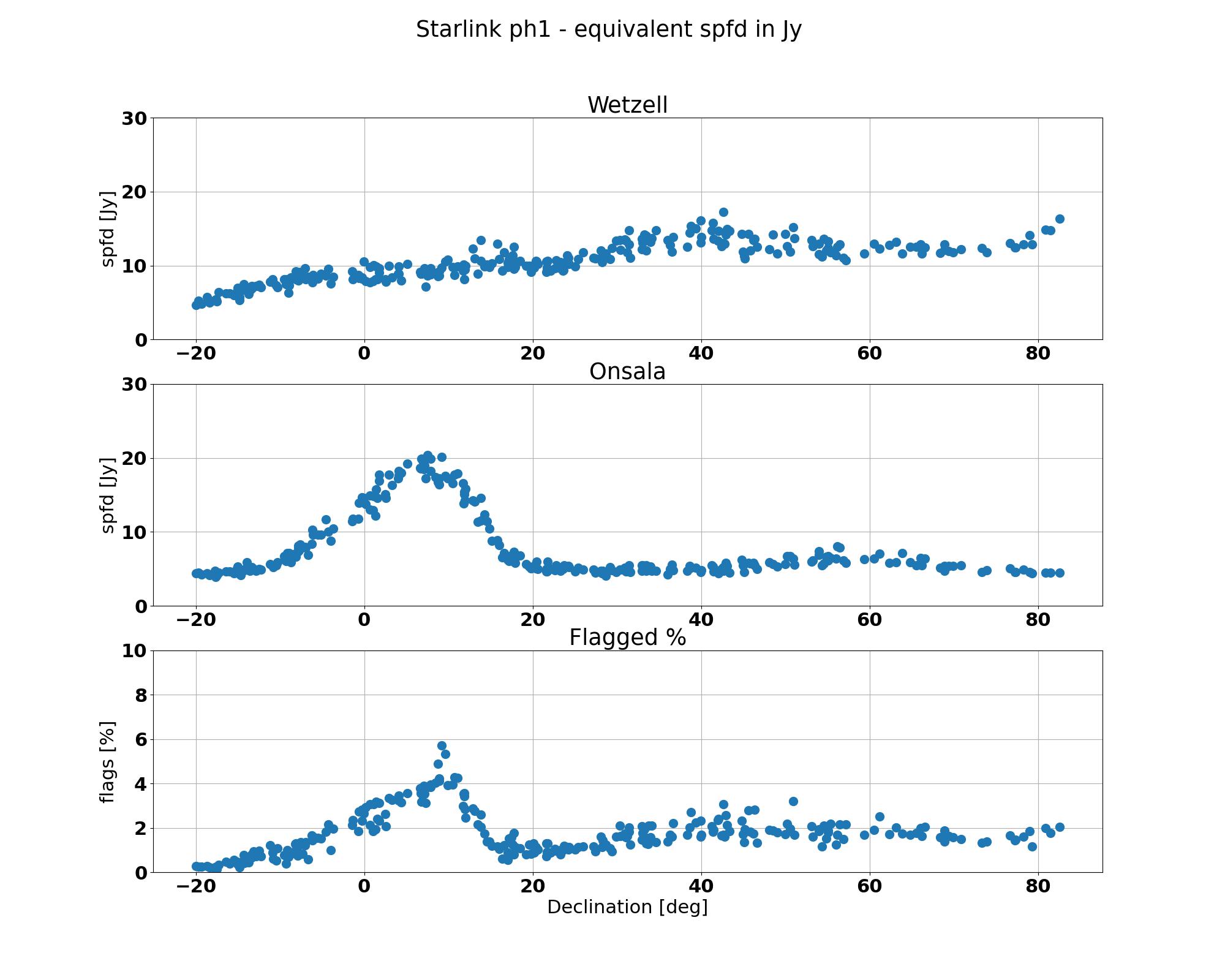}
\includegraphics[width=7.8cm]{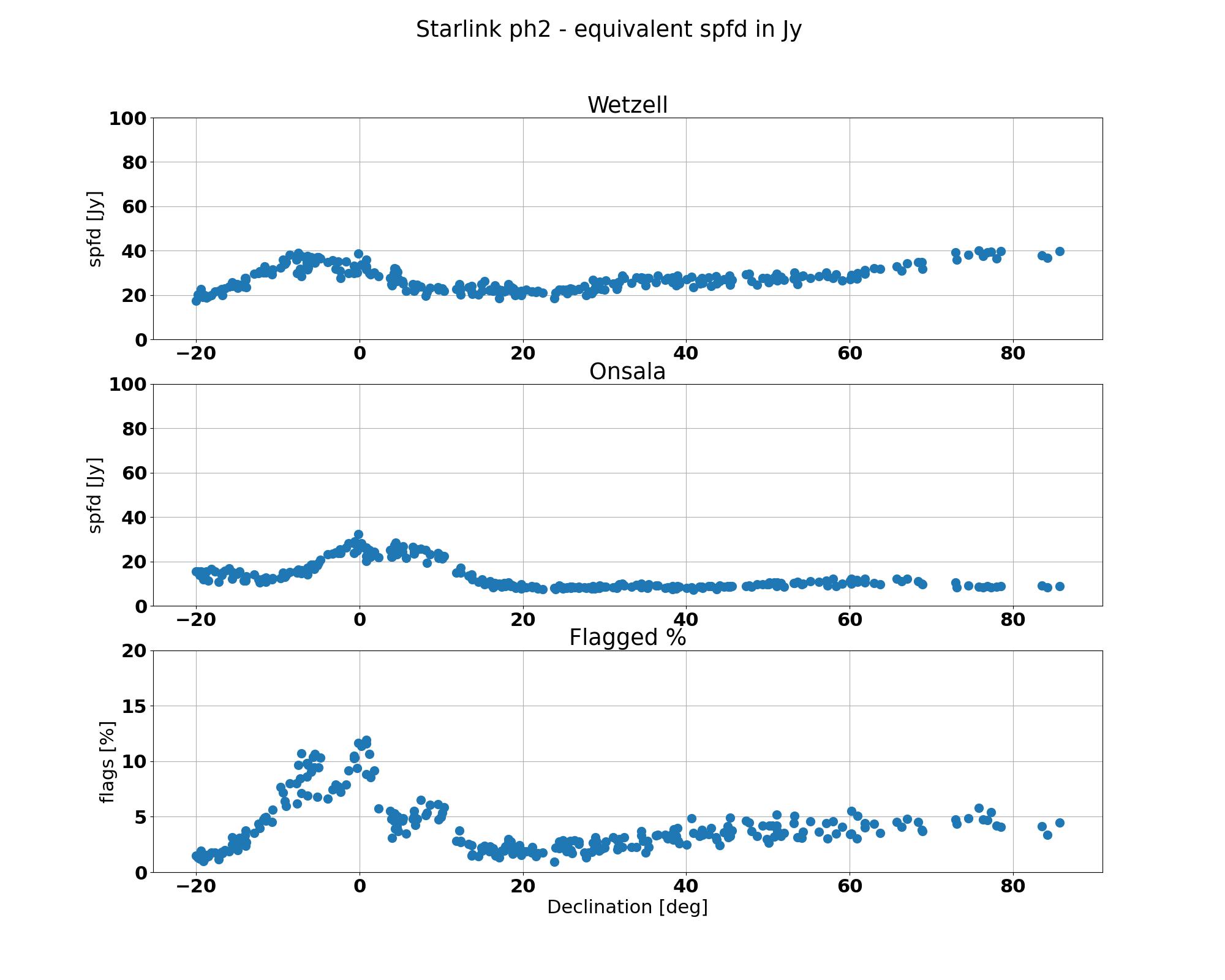}
\caption{Average Equivalent Spectral Power Flux Density (aESPFD) as a function of declination for each constellation.}
\label{aESPFD}
\end{figure}

Considering that the ITU-R RA.769 thresholds for harmful interference for VLBI are defined as $-193~ dBW/m^2/Hz$, representing an ESPDF of $250~Jy$ in an antenna of $13~m$ diameter, the results show that VGOS observations could in principle be conducted inside the satellite downlink bands (considering the percentage of data lost). See Figure \ref{aESPFD}.

\section{Conclusions}
This paper proposed metrics to evaluate the impact of large satellite constellations on VGOS operations by a simil-epfd simulation for Starlink ph1 and ph2, and  OneWeb ph1, and two European stations as receivers. 

Through calculations and simulations it was proved that the maximum received power even in beam-to-beam coupling condition with satellites will not be enough to saturate the analog chain of a VGOS receiver.
As for the digitized part, the simulations show that observations in the same band as the downlinks from satellites can have a significant percentage of data loss due to strong signals compared to the thermal noise of the receiver. Nevertheless the results shows that the ESPFD for both antennas and all constellations is lower than the thresholds defined by ITU-R for VLBI. Observations outside of the satellite downlink bands should not be impacted by satellite downliks in this frequency range.

As further work the authors will continue investigating how correlation can help mitigate this signals from satellite constellations and how the aggregation of all constellations scales the impact.

\section*{Acknowledgements}

The authors would like to thank the IVS Coordinating Center at NASA Goddard Space Flight Center (GSFC) for taking the archive of IVS sessions. The schedule used in this work is available at the https://ivscc.gsfc.nasa.gov/sessions/2022/vo2027 web page. We are grateful to Salvo Buttaccio, for the assistance with the VGOS schedule, to Dr. Benjamin Winkel for assistance with the use of the Cysgp4 Python package, and to Dr. Jose Antonio Lopez-Perez and Dr. Hayo Hase for useful discussions about VGOS receivers and operations.


\begin{thebibliography}{99}

\bibitem{RFI_Baan}
W. A. Baan, 2011. "RFI mitigation in radio astronomy"\\
RFI Mitigation Workshop 2010

\bibitem{Cohen}
J. Cohen, Iridium and Radio Astronomy in Europe\\
Spectrum Management for Radio Astronomy: proceedings of the IUCAF summer school held at Green Bank, West Virginia, June 9-14, 2002.

\bibitem{Cooper_bits}
Cooper, B.F.C., 1970. "Correlators with two-bit quantization".\\ Australian Journal of Physics, 23, pp.521-527.

\bibitem{ECC271}
ECC Report 271, "Compatibility and sharing studies related to NGSO satellite systems operating in the FSS bands 10.7-12.75 GHz (space-to-Earth) and 14-14.5 GHz (Earth-to-space)"\\
European Communications Office, 2021

\bibitem{Lawrence}
A. Lawrence Et. Al., "The case for space environmentalism"
Nature Astronomy volume 6, pages428–435 (2022)

\bibitem{OneWeb_ph1}
OneWeb phase 1 FCC filing \\
\url{https://fcc.report/IBFS/SAT-MPL-20200526-00062/2379565}

\bibitem{Petrachenko_RFI}
B. Petrachenko, "The Impact of Radio Frequency Interference (RFI) on VLBI2010" \\
IVS 2010 General Meeting Proceedings, p.434–438

\bibitem{Petrachenko_WG3}
B. Petrachenko et. al. 2010. "Final Report of the Observing Strategies Sub group of the IVS Working Group 3"\\
IVS 2010 General Meeting\\
\url{https://ivscc.gsfc.nasa.gov/about/wg/wg3/1\_observing\_strategies.pdf}

\bibitem{RA.769}
RECOMMENDATION ITU-R RA.769 "Protection criteria used for radio astronomical measurements"

\bibitem{RA.1631}
RECOMMENDATION ITU-R RA.1631 "Reference radio astronomy antenna pattern to be used for compatibility analyses between non-GSO systems and radio astronomy service stations based on the epfd concept"

\bibitem{S.1586}
RECOMMENDATION ITU-R S.1586 "Calculation of unwanted emission levels produced by a non-geostationary fixed-satellite service system at radio astronomy sites"

\bibitem{Starlink_ph1}
Starlink phase 1 FCC filing
\url{https://fcc.report/IBFS/SAT-MOD-20200417-00037/2274316}

\bibitem{Starlink_ph2}
Starlink phase 2 FCC filing
\url{https://fcc.report/IBFS/SAT-AMD-20210818-00105}

\bibitem{Astropy}
The Astropy Collaboration et.al., "Astropy: A community Python package for astronomy"
A\&A Volume 558, October 2013

\bibitem{Astropy2}
The Astropy Collaboration et.al.,"The Astropy Project: Building an inclusive, open-science project and status of the v2.0 core package"
\url{https://arxiv.org/abs/1801.02634}

\bibitem{Thompson_RFI}
Thompson, 1982. "The Response of a Radio-Astronomy Synthesis Array to Interfering Signals"
IEEE TRANSACTIONS ON ANTENNAS AND PROPAGATION, VOL. AP-30, NO. 3, MAY 1982

\bibitem{Walker}
J. G. Walker, Satellite constellations, Journal of the British Interplanetary Society, vol. 37, pp. 559-571, 1984

\bibitem{Cysgp4}
B. Winkel, "A wrapper around the SGP4 package, for sat TLE calculations"
\url{https://github.com/bwinkel/cysgp4}

\end{thebibliography}
\end{document}